\documentclass[reqno]{article}
\usepackage{alltt}
\usepackage{graphicx}
\title{Particle - unparticle duality in super-relativity}
\author{Peter Leifer}
\date{Cathedra of Informatics, Crimea State Engineering and
Pedagogical University, \\
21 Sevastopolskaya st., 95015 Simferopol, Crimea, Ukraine; \\
leifer@bezeqint.net }
\begin{document}
\maketitle
\begin{abstract}
New method of shaping quantum ``particle - unparticle" vacuum excitations has been proposed
in the framework of unification of relativity and quantum theory. Such unification
is based solely on the notion of generalized coherent state (GCS) of N-level system
and the geometry of unitary group $SU(N)$ acting in state space $C^N$. Initially,
neither contradictable notion of quantum particle, nor space-time coordinates
(that cannot be a priori attached
to nothing) are used in this construction. Quantum measurement of local
dynamical variables (LDV) leads to the emergence of $4D$ dynamical space-time (DST).
Morphogenesis of the ``field shell" of GCS and its dynamics have been studied
for $N=2$ in DST.
\end{abstract}
\vskip 0.1cm
\section{Introduction}
Statistical analysis of the energy distribution is the base of the black body
radiation \cite{Planck1} and the Einstein's theory of the light emission and absorption
\cite{Einstein_Q}. Success of Einstein hypothesis of photons, de Broglie wave concept
of particles \cite{dB} and Schr\"odinger's equation for hydrogen atom \cite{Sch1} paved
the way to corpuscular-wave duality of matter.
This conceptual line was logically finished by Dirac in his
method of the second quantization \cite{Dirac1}. This approach is perfectly fits to
many-body weakly interacting quantum systems and it was assumed that the
``corpuscule-wave duality" is universal. However the application of this method to
single quantum ``elementary" particles destroys this harmony. Physically it is clear
why: quantum particle is self-interacting system and this interaction is at least of the
order of its rest mass. Since the nature of the mass is the open problem we do not know
the energy distribution in quantum particles up to now. Here I try to show a possible
approach to this problem in the framework of simple model in pure deductive manner.

A long time it is was assumed
that the dynamical model may be found in the framework of the string theory, but the
epitaph to string theory \cite{Schroer} clearly shows the deep crisis of particle
physics in its present form. Notice, Einstein \cite{EPR} and
Schr\"odinger \cite{Schr} treated the statistical fundament of quantum theory as a
perishable and temporal. Quantum theory solved a lot of fundamental problems, but
(as it happen with fundamental theory), it posed number of deeper questions.
Even first steps in wave picture of quantum particles brought sudden surprises.
First of all there was a big discrepancy between intuitive Schr\"odinger's imagination
and real properties of ``corpuscular waves".
Initially Schr\"odinger thought that there is a possibility to build stable wave packet
from plane waves of de Broglie that may be treated as the wave model of localized
electron; he understood soon that it is impossible. Only in some special
case of quantized harmonic oscillator he could build such stable wave packet moving
(in average) like material point under Hooke's elastic force \cite{Sch2}.

Historically, the impossibility to get wave description of localizable particle
led to probabilistic interpretation of the wave function. In fact, this is  the fork point
changing all character of fundamental physics: state vector is treated as amplitude of
probability for particle to be in some particular state. This paradigm is the
source of all fundamental unsolved problems mentioned above: measurement problem,
localization, divergences, etc. However, practical applications of quantum theory are so
convincible and prolific that any attempts to find answers on ``children questions"
are frequently treated as a pathology. Nevertheless, we should analyze these problems
again and again till all situation will be absolutely clear without any references to
the beauty of quantum mystic \cite{Feynman1}. Two fundamental problems
should be solved: how to get (from first principles) non-linear quantum field equations
with localizable solutions and how to formulate objective quantum measurement of
dynamical variables?

Dirac's equations for electron in Coulomb
potential of nuclei are realistic in first approximation but higher approximations
suffer from divergences. Expressions for self-energy, electric charge and
magnetic momentum of self-interacting electron demonstrate divergences too.
Renormalization procedure is in fact the attempt to correct this construction.
The same equations being applied to ``free electron" have plane
wave solutions that never where observed and that may be rather related to co-vector
generated by periodic lattice \cite{G}.  Physical interpretation of plane wave solution
being applied to single free particle requires essential efforts. Furthermore, the
price of these efforts is unacceptably high: probabilistic interpretation,
collapse of wave function, parallel worlds, Multiverse etc., are invoked to explain the
formal, unobservable and artificial solution.

Plane waves and $\delta$-function are examples of improper states that may be formally
incorporated in Hilbert manifolds on equal footing with square-integrable states
in the framework of ``functional relativity" \cite{Kryukov1,Kryukov2}.
Thereby, the classical ideal notion of pointwise interaction is deemed as legitimized.
I think, however, that such approach is acceptable as effective temporal method when 
discussion in nature of quantum interaction is postponed.

Generally, improper states like $\delta$-function and plane waves are only artifact
arose as assumption of applicability of our fundamental linear equations
for ``free" quantum particle. In fact, the principles of Hamilton or
Lagrange can not fundamentally to determine the quantum dynamics
because the concept of quantum amplitudes
shows that these principles (being formulated as the principle
of the least action) are merely (sometimes bad mathematically defined) \emph{approximation}.
One should found some more general quantum principle of
self-organization (morphogenesis) and dynamics of quantum matter.

Few years ago I derived field quasi-linear PDE as a consequence
of conservation law of local Hamiltonian on quantum phase space $CP(N-1)$ \cite{Le6,Le7}.
This conservation law was expressed as affine parallel transport of
Hamiltonian vector field in $CP(N-1)$ in connection agreed with Fubini-Strudy
metric \cite{KN,Le2,Le6,Le7}. These quasi-linear PDE have soliton-like solutions
whose physical status is unknown. New investigations in so-called ``unparticle"
area \cite{Georgi} gave me some hint on possible interpretation of gotten equations.

I would like to discuss here a morphogenesis of quantum particle in the spirit of reaction
$e^-\to \mathcal{U} \to e^-$. In other words I propose to study the particle/unparticle
sectors of matter \cite{Georgi} in wide range of energy in order to solve localization
problem in foundations of quantum physics. The concept of \emph{scale invariance}
\cite{Georgi,Yuan} will be replaced by the principle of super-relativity \cite{Le1,Le2}.

I should note that Blochintzev about 60 years ago discussed the unparticle
sector in the framework of universality
of wave - particle ``duality" for interacting quantum fields  \cite{Bl1,Bl2}.
For such fields the universality is generally broken.
Namely, attempt to represent two interacting boson fields as the set of free quantum
oscillators leads to two types of oscillators: quantized and non-quantized. The second
one arises under simple relation $g > \frac{m_1 m_2c^2}{h^2}$ between coupling constant $g$
and masses $m_1$ and $m_2$ of two scalar fields. For such intensity of coupling we obtain a
field with excitation states in two sectors: particle and ``unparticle". Furthermore,
the excitations in ``unparticle" sector has an imaginary mass and they propagate with group
velocity larger than $c$. For self-interacting scalar field
of mass $m$ the intensity of self-interaction $g$ leads to breakdown of the universality
of the wave - particle ``duality" if it is larger than the inverse square of the Compton
wavelength: $g > \frac{m^2c^2}{h^2}=\frac{1}{\lambda^2_C}$.

\section{The Action State Space}
Blochintzev's examples were oversimplified for clarity.
We have to have the process of morphogenesis of quantum particle/unparticle sectors
that should be dynamically described.
One may even think that interacting observable particles are
immersed into the sea of ``unparticle" excitations somehow related
with ``dark matter". It leads to necessity to modify
the second quantization method. Besides arguments of Blochintzev
there at least two reasons for such modification.

{\it First.} In the second quantization method one has formally
given particles whose properties are defined by some commutation
relations between creation-annihilation operators. Note, that the
commutation relations are only the simplest consequence of the
curvature of the dynamical group manifold in the vicinity of the
group's unit (in algebra). Dynamical processes require, however,
finite group transformations and, hence, the global group structure.
The main my technical idea is to use vector fields over a group
manifold instead of Dirac's abstract q-numbers. This scheme
therefore seeks the dynamical nature of the creation and
annihilation processes of quantum particles.

{\it Second.} The quantum particles (energy bundles) should
gravitate. Hence, strictly speaking, their behavior cannot be
described as a linear superposition. Therefore the ordinary second
quantization method (creation-annihilation of free particles) is
merely a good approximate scheme due to the weakness of gravity.
Thereby the creation and annihilation of particles are time
consuming dynamical non-linear processes. So, linear operators of
creation and annihilation (in Dirac sense) do exist as approximate
quantities.

POSTULATE 1.

\noindent {\it There are elementary quantum states $|\hbar a>,
a=0,1,...$ belonging to the Fock space of an abstract Planck
oscillator whose states correspond to the quantum motions with given
number of Planck action quanta}.

One may image some {\it ``elementary quantum states"
(EAS) $|\hbar a>$ as a quantum motions with entire number $a$ of the
action quanta}. These $a,b,c,...$ takes the place of the ``principle
quantum number" serving as discrete indices $0 \leq a,b,c... <~
\infty$. Since the action by itself does not create gravity, but only velocity
of action variation, i.e. energy/matter, it is
possible to create the linear superposition of $|\hbar
a>=(a!)^{-1/2} ({\hat \eta^+})^a|\hbar 0>$ constituting $SU(\infty)$
multiplete of the Planck's action quanta operator $\hat{S}=\hbar
{\hat \eta^+} {\hat \eta}$ with the spectrum $S_a=\hbar a$ in the
separable Hilbert space $\cal{H}$. Therefore, we shall primarily
quantize the action, not the energy. The relative (local) vacuum
of some problem is not necessarily the state with minimal energy, it
is a state with an extremal of some action functional.

The space-time representation of these states and their coherent
superposition is postponed on the dynamical stage as it is described
below. We shall construct non-linear field equations describing
energy (frequency) distribution between EAS's $|\hbar a>$, whose
soliton-like solution provides the quantization of the dynamical
variables. Presumably, the stationary processes are represented by
stable particles and quasi-stationary processes are represented by
unstable resonances or unparticle stuff.

Generally the coherent superposition
\begin{eqnarray}
|F>=\sum_{a=0}^{\infty} f^a| \hbar a>,
\end{eqnarray}
may represent of a ground state or a ``vacuum" of some quantum
system with the action operator
\begin{eqnarray}
\hat{S}=\hbar A({\hat \eta^+} {\hat \eta}).
\end{eqnarray}
Then one can define the action functional
\begin{eqnarray}
S[|F>]=\frac{<F|\hat{S}|F>}{<F|F>},
\end{eqnarray}
which has the eigen-value $S[|\hbar a>]=\hbar a$ on the eigen-vector
$|\hbar a>$ of the operator $\hbar A({\hat \eta^+} {\hat
\eta})=\hbar {\hat \eta^+} {\hat \eta}$ and that deviates in general
from this value on superposed states $|F>$ and of course under a
different choice of $\hat{S}=\hbar A({\hat \eta^+} {\hat \eta}) \neq
\hbar {\hat \eta^+} {\hat \eta}$. In order to study the variation of
the action functional on superposed states one need more details on
geometry of their superposition.

In fact only finite, say, $N$ elementary quantum states (EQS's)
($|\hbar 0>, |\hbar 1>,...,|\hbar (N-1)>$) may be involved in the
coherent superposition $|F>$. Then $\cal{H}=C^N$ and the ray space
$CP(\infty)$ will be reduced to finite dimensional $CP(N-1)$.
Hereafter we will use the indices as follows: $0\leq a,b \leq N-1$,
and $1\leq i,k,m,n,s \leq N-1$. This superposition physically
corresponds to the complete amplitude of quantum motion in setup $S$.
Then GCS corresponding to this amplitude is controlled by $SU(N)$
dynamical group. One may assume that following postulate takes the place:

POSTULATE 2.

\noindent {\it Matter (energy) distribution is determined by velocities of GCS variations
by LDV's like local Hamiltonian}.

Realization of this assumption will be discussed below.

\section{From flexible setup to quantum reference frame in super-relativity}
Let me show how ordinary quantum formalism hints us how to formulate
functionally invariant quantum dynamics.

\subsection{Flexible setup in the action state space}
The ordinary quantum formalism of operations with amplitudes was brightly
demonstrated by Feynman in popular lectures \cite{Feynman1}. This formalism shows
that generally two setups $S_1$ and $S_2$ lead to
different amplitudes $|\Psi_1>, |\Psi_2>$ of outcome event. There are infinite number
of different setups and not only in the sense of different space-time position but
in different parameters of fields, using devices, etc. Symmetries
relative space-time transformations of whole setup have been studied in ordinary
quantum approach. Such approach reflects, say, the
\emph{first order of relativity}:
the physics is same if any \emph{complete setup} subject (kinematical, not
dynamical!) shifts, rotations, boosts as whole in Minkowski space-time.

Next step leading to new type of relativity may be formulated as
invariance of physical properties of quantum particles
lurked behind two amplitudes $|\Psi_1>, |\Psi_2>$.
Similar idea in the framework of ``functional relativity" was formulated by A. Kryukov
\cite{Kryukov1,Kryukov2} as a requirement that before and after interaction the wave
function of electron should have functionally invariant form. I will treat this
requirement
as ``global functional relativity" since the process of transition from ``in"-state to
``out"-state is left outside of envision. It is shown by clear example of ``interaction"
with a spectrometer. In this ad hoc taken measurement all root problems are hidden in
two assumptions:

1. classical motion of pointwise electron in spectrometer, and

2. in pointwise absorption of the electron by screen.

These simplifications gave a possibility to treat the inverse Fourier transform
as the spectrometer action and to use the Gaussian kernel
$k_{\tilde{H}}(y,v)=e^{-\frac{1}{2}(y-v)^2}$ playing the role of metric in Hilbert space
of improper states. The question however is: what happen in more general kind of
interaction where electron is participated? Is it possible to build the mathematical
model of interaction in relativistic case, say, for high energy reactions like
$e^- e^+ \to \gamma + \gamma \to \tau^- \tau^+$? Definitely, this problem could not
formulated in the spirit of ``global functional relativity". In order to describe
smooth quantum evolution let me ask: what happen if I slightly variate some device
in the setup, say, rotate a filter or, better, change magnetic field around dense flint
\cite{Le4} in complete setup? In other words I will use
``local functional relativity" or ``super-relativity" by declaration that infinitesimal
variation of setup by small variations of its parameters leads to small variations of
output state. Now not space-time coordinates
play essential role but some internal parameters like strength of field used in given setup.
 But how we should formalize
``physics" and its invariance mathematically? Our model should be maximally simple
since we would like to study very basic properties of quantum physics.
There is a fine technical question about parametrization of output state
as function of fields in devices, adjustments, etc. It would be a mistake
to start our description from ``given" particles in space-time and fields of setup
since neither particle nor space-time are good enough defined at this stage.
Any real physical setup is even much more
complicated system and its classical parametrization is, however,
unacceptable for our aim since it returns us to Borh's tenet
that all quantum relations should be expressed ``by classical language".
Then we will be involved in the routine round of quantum measurement problem.

The key step to the
invariant description of quantum state $|S>=\sum_{a=0}^{N-1}S^a|\hbar a>$ is transition
to local functional coordinates $\pi^i_{(j)}=\frac{S^i}{S^j}$ of its GCS in $CP(N-1)$
that carrying representation of $SU(N)$ dynamical group \cite{Le1,Le2,Le8}. Now local
quantum reference frame parameterizations by local functional coordinates $\pi^i$
of GCS should be used. Here arises the \emph{second order of relativity which I
called super-relativity: the physics of some quantum object corresponding to GCS
of $|S>$ is same in any setup}.

\subsection{Super-relativity}
The principle of super-relativity arose as development of
Fock's idea of ``relativity to measuring device" \cite{Fock}. This idea may be
treated as generalization of the relativity principle in space-time to
``functional relativity" in the state space \cite{Kryukov1,Kryukov2}
under some reservations and specification. However the power of Fock's program is
limited in comparison with power of Einstein's concepts of special and general relativity.
The main reason is that the notion of the ``measuring device" could not be
correctly formulated in the own framework of the standard quantum theory. Some
additional and, in fact, outlandish classical ingredients should be involved.
Same argument may be applied to the ``global functional relativity" since only
in some particular case it is possible to find theoretically analyzable model
of quantum setup comprising classical improper states (like plane waves and
$\delta$-function) as it was discussed above.
In order to overcome this problem we should to clarify relations between
state vector and dynamical variables of quantum system.

It is very strange to think that state vector being treated as basic element
of the \emph{full} description of quantum system does not influence on dynamical
variable of quantum system. Ordinary quantum dynamical variables are represented by
hermitian operator in Hilbert space carrying representation of symmetry group,
say, Lorentz group. All formal apparatus of quantum theory is based on the
assumption that operators of position, momentum, etc. depend only upon the parameters of
Lorentz group.
Lets now assume that we would like to investigate general behavior of quantum state
vector $|S>$ in Hilbert space $\mathcal{H}=C^N$ subject control
by unitary group $SU(N)$ \cite{Le1,Le2,Le3} through group parameters
$\Omega^{\alpha}: \quad 1 \leq \alpha \leq N^2-1$. This state may be represented as
$|S>=\sum_{a=0}^{N-1}S^a|\hbar a>$ where space-time coordinates are not even mentioned
and whose dynamics and ``morphogenesis" somehow
related to space-time coordinates which I will discuss later.
I argue that this approach:

1. does not require any classical model,

2. it represents due to its generality the $SU(N)$ in dynamical
space-time (inverse representation) through the ``morphogenesis" of the
``field shell",

3. ``field shell" of GCS obeys to quasi-linear PDE in dynamical space-time that
may me solved in some reasonable approximation.
The pure local in quantum state space $CP(N-1)$ theory uses the local geometry of
$SU(N)$ group. The group parameters takes the place of non-Abelian gauge fields
surrounding quantum object (particle/unparticle) whose properties a priori are unknown.
But small variations of these fields lead to small variation
of GCS that and may be associated with some state-dependent LDV modeling flexible setup
or quantum reference frame (QRF).

In order to keep the invariant
properties of some quantum particle (probably, better to say ``quantum process" since
particle may or may not arise during this process) involved in this manifold of
setups one should know \emph{difference in amplitudes arose due to setup variation}.
\emph{Most technically important approach is the comparison of dynamical variables at
infinitesimally close quantum states arose in slightly different setups.}
Since stationary states may be represented by rays in Hilbert space, I will work
in projective Hilbert state space $CP(N-1)$. This approach leads to the concept
of LDV  taking the place of quantum reference frame and to the principle of
super-relativity \cite{Le1,Le2} that may be expressed as follows:

POSTULATE 3.

\noindent {\it Unitary transformations
$U(\tau)=\exp(i\Omega^{\alpha}\hat{\Lambda}_{\alpha}\tau)$ of the action amplitudes
may be identified with physical fields. Field functions $\Omega^{\alpha}$ are in the
adjoint representation of $SU(N)$, $\hat{\Lambda}_{\alpha} \in AlgSU(N)$, and $\tau$
is an evolution parameter. The coset transformations $G/H=SU(N)/S[U(1)\times U(N-1)]=CP(N-1)$
is the quantum analog of  classical force; its action is equivalent to physically
distinguishable deformation of GCS in $CP(N-1)$, isotropy group $H=U(1)\times U(N-1)$
takes the place of the of gauge group}.

Since any state $|S>$ has the isotropy group
$H=U(1)\times U(N)$, only the coset transformations $G/H=SU(N)/S[U(1)
\times U(N-1)]=CP(N-1)$ effectively act in $C^N$. One should remember,
however, that the Cartan decomposition of unitary group has the
physical sense only in respect with initially chosen state vector. Therefore the
parametrization of these decomposition is state-dependent
$[h_{|S>},h_{|S>}] \subseteq h_{|S>}, [b_{|S>},b_{|S>}]
\subseteq h_{|S>}, [b_{|S>},h_{|S>}] \subseteq b_{|S>}$ \cite{Le1,Le2,Le3}.
It means that physically it is interesting not abstract unitary group relations
but realization of the unitary group transformations resulting in motion of the pure
quantum states represented by rays in projective Hilbert space. Therefore the
ray representation of $SU(N)$ in $C^N$, in particular, the embedding
of $H$ and $G/H$ in $G$, is a state-dependent parametrization.
This is a key point of all construction
invoking to life the concept of the LDV expressed by tangent
vectors fields to $CP(N-1)$. Technically it means that the local $SU(N)$ unitary
classification of the quantum motions of GCS and distinction between particles
and unparticles requires
the transition from the matrices of Pauli $\hat{\sigma}_{\alpha},(\alpha=1,...,3)$,
Gell-Mann $\hat{\lambda}_{\alpha},(\alpha=1,...,8)$, and in general $N \times N$ matrices
$\hat{\Lambda}_{\alpha}(N),(\alpha=1,...,N^2-1)$ of $AlgSU(N)$ to the tangent vector
fields to $CP(N-1)$ in local coordinates \cite{Le1}.
Hence, there is a diffeomorphism between the space of the rays
marked by the local coordinates
\begin{equation}
\pi^i_{(j)}=\cases{\frac{S^i}{S^j},&if $ 1 \leq i < j$ \cr
\frac{S^{i+1}}{S^j}&if $j \leq i < N-1$}
\end{equation}\label{coor}
in the map
 $U_j:\{|S>,|S^j| \neq 0 \}, j\geq 0$
and the group manifold of the coset transformations
$G/H=SU(N)/S[U(1) \times U(N-1)]=CP(N-1)$.
This diffeomorphism is provided by the coefficient functions
\begin{equation}\label{17}
\Phi_{\sigma}^i = \lim_{\epsilon \to 0} \epsilon^{-1}
\biggl\{\frac{[\exp(i\epsilon \hat{\Lambda}_{\sigma})]_m^i S^m}{[\exp(i
\epsilon \hat{\Lambda}_{\sigma})]_m^j S^m }-\frac{S^i}{S^j} \biggr\}=
\lim_{\epsilon \to 0} \epsilon^{-1} \{ \pi^i(\epsilon
\hat{\Lambda}_{\sigma}) -\pi^i \}
\end{equation}
of the local generators
\begin{equation}\label{18}
\overrightarrow{D}_{\sigma}=\Phi_{\sigma}^i \frac{\partial}{\partial \pi^i} + c.c.
\end{equation}
comprise of non-holonomic overloaded basis of $CP(N-1)$ \cite{Le1}.
Here $\epsilon$ is used for one of the $SU(N)$ parameters $\Omega^{\sigma}$.
Now one may
introduce local Hamiltonian as a tangent vector fields
\begin{equation}\label{19}
\overrightarrow{H}=\hbar \sum_{\sigma = 1}^{N^2 -1}\Omega^{\sigma}(\tau)
\overrightarrow{D}_{\sigma}=\hbar
\sum_{\sigma = 1}^{N^2-1}\Omega^{\sigma}(\tau)\Phi_{\sigma}^i \frac{\partial}{\partial
\pi^i} + c.c.
\end{equation}
whose coefficient functions $\Omega^{\sigma}(\tau)$ may be found under the condition of
self-conservation expressed as affine parallel transport of Hamiltonian vector field
$H^i=\Omega^{\sigma}(\tau)\Phi_{\sigma}^i$ agrees with Fubini-Study metric.
The problem of finding $\Omega^{\sigma}(\tau)$ treated in the
context of gauge field application as surrounding fields of quantum lump was discussed
in \cite{Le5,Le6}.

The ``visualization" of this gauge field requires the attachment
of co-movable ``Lorentz frame" in DST. I use the analogy with clock's arm
shows the Abelian phase of wave function (see Figure 1) in Feynman's simplified
explanation of quantum electrodynamics \cite{Feynman1}.  Feynman discussed the
\emph{amplitude of an event in stationary situation} since operations with
amplitudes refer to fixed setup.
\begin{figure}[h]
  \includegraphics[width=1in]{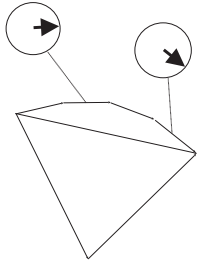}\\
  \caption{Fixed setup - summation and multiplication of state vectors
  of being particles. Feynman's summation of amplitudes corresponding
  to the time of light propagation from internal points of glass plate
  to detector. Equivalent amplitude arises as sum of ``forward" and ``backward"
  reflection from border surfaces.}\label{fig.1}
\end{figure}
Dynamical GCS moving due to $\Omega^{\sigma}(\tau)$ variation requires operation with
velocities of
state deformation. This variable setup is described by LDV wrapped into ``field shell" that
should dynamically conserve local Hamiltomian vector field \cite{Le2,Le3,Le4}.
I attached qubit spinor and further ``Lorentz frame" that define ``4-velocity" of
some imaging point (belonging to the DST) of the quantum dynamics.  This imaging point
is the mentioned above analog of clock's arrow but now in $4D$ DST, see Figure 2.
Quasi-linear partial differential equations arising as a consequence of conservation
law of local Hamiltonian of evolving quantum system, define morphogenesis of non-Abelian
(phase) gauge soliton-like ``field shell" \cite{Le5,Le6,Le8}. So, we have a concentrated
``lump" associated with becoming quantum particle.
\begin{figure}[h]
  \includegraphics[width=1in]{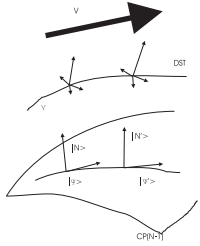}\\
  \caption{Dynamical setup for becoming lump - operations with LDV.
  In order to get effective sum of non-Abelian phases of $SU(N)$ transformation
  shaping the lump, one should integrate quasi-linear partial differential
  equations. The ``4-velocity" $V$ of imaging point in DST is parameterized by
  boosts and angle velocities of co-moving ``Lorentz reference frame" attached
  to trajectory in $CP(N-1)$.}\label{fig.2}
\end{figure}
Such particle may be represented as a dynamical process due to morphogenesis
of the ``field shell" of generalized coherent state of N-level system.

\section{Local dynamical variables}
The action state space ${\cal H}=C^N$
contains ``initial" and ``final" stationary states with finite action quanta.
Quantum dynamics is described by {\it the
velocities of the GCS variation} representing some ``elementary
excitations'' (quantum particles or unparticles). Their dynamics is specified by
the Hamiltonian, giving time variation velocities of the action
quantum numbers in different directions of the tangent Hilbert space
$T_{(\pi^1,...,\pi^{N-1})} CP(N-1)$ which takes the place of the
ordinary linear quantum scheme as will be explained below. The rate
of the action variation gives the energy of the excitations in accordance
with POSTULATE 2.

The local dynamical variables correspond to internal symmetries of
the GCS and their evolution should be expressed now in terms of the
local coordinates $\pi^k$. The Fubini-Study metric
\begin{equation}
G_{ik^*} = [(1+ \sum |\pi^s|^2) \delta_{ik}- \pi^{i^*} \pi^k](1+
\sum |\pi^s|^2)^{-2} \label{FS}
\end{equation}
and the affine connection
\begin{eqnarray}
\Gamma^i_{mn} = \frac{1}{2}G^{ip^*} (\frac{\partial
G_{mp^*}}{\partial \pi^n} + \frac{\partial G_{p^*n}}{\partial
\pi^m}) = -  \frac{\delta^i_m \pi^{n^*} + \delta^i_n \pi^{m^*}}{1+
\sum |\pi^s|^2} \label{Gamma}
\end{eqnarray}
in these coordinates will be used. Hence the internal dynamical
variables and their norms should be state-dependent, i.e. local in
the state space \cite{Le1,Le2}. These local dynamical variables
realize a non-linear representation of the unitary global $SU(N)$
group in the Hilbert state space $C^N$. Namely, $N^2-1$ generators
of $G = SU(N)$ may be divided in accordance with the Cartan
decomposition. There are
$(N-1)^2$ generators
\begin{eqnarray}
\Phi_h^i \frac{\partial}{\partial \pi^i}+c.c. \in H,\quad 1 \le h
\le (N-1)^2
\end{eqnarray}
of the isotropy group $H = U(1)\times U(N-1)$ of the ray (Cartan
sub-algebra) and $2(N-1)$ generators
\begin{eqnarray}
\Phi_b^i \frac{\partial}{\partial \pi^i} + c.c. \in B, \quad 1 \le b
\le 2(N-1)
\end{eqnarray}
are the coset $G/H = SU(N)/S[U(1) \times U(N-1)]$ generators
realizing the breakdown of the $G = SU(N)$ symmetry of the GCS.
Furthermore, the $(N-1)^2$ generators of the Cartan sub-algebra may
be divided into the two sets of operators: $1 \le c \le N-1$ ($N-1$
is the rank of $Alg SU(N)$) Abelian operators, and $1 \le q \le
(N-1)(N-2)$ non-Abelian operators corresponding to the
non-commutative part of the Cartan sub-algebra of the isotropy
(gauge) group. Here $\Phi^i_{\sigma}, \quad 1 \le \sigma \le N^2-1 $
are the coefficient functions of the generators of the non-linear
$SU(N)$ realization. They give the infinitesimal shift of the
$i$-component of the coherent state driven by the $\sigma$-component
of the unitary multipole field $\Omega^{\sigma}$ rotating the
generators of $Alg SU(N)$ and they are defined as by (5)
\cite{Le1,Le2}. Then the sum of the $N^2-1$ the energies associated with
intensity of deformations of the GCS is represented  by the local
Hamiltonian vector field $\vec{H}$ which is linear in the partial
derivatives $\frac{\partial }{\partial \pi^i} = \frac{1}{2}
(\frac{\partial }{\partial \Re{\pi^i}} - i \frac{\partial }{\partial
\Im{\pi^i}})$ and $\frac{\partial }{\partial \pi^{*i}} = \frac{1}{2}
(\frac{\partial }{\partial \Re{\pi^i}} + i \frac{\partial }{\partial
\Im{\pi^i}})$. In other words it is the tangent vector to $CP(N-1)$
\begin{eqnarray}
\vec{H}=\hbar \Omega^c \Phi_c^i \frac{\partial }{\partial
\pi^i} + \hbar \Omega^q \Phi_q^i \frac{\partial }{\partial \pi^i} +
\hbar \Omega^b \Phi_b^i \frac{\partial }{\partial \pi^i} + c.c.
\label{field}
\end{eqnarray}
Thereby in the framework of the local state-dependent
approach one can formulate a quantum
scheme with help more flexible mathematical structure
than  matrix formalism. It means that matrix elements of
transitions between {\it two arbitrary far states} are
associated with, in fact, bi-local dynamical
variables that
bring a lot of technical problems in quantum field area.
However the local dynamical
variables related to infinitesimal deformations of quantum
states are well defined in projective Hilbert
space as well as quantum states itself.
They are local tangent vector fields to the
projective Hilbert space $CP(N-1)$ which
are $SU(N)$ generators (differential operators
of first order) \cite{Le1,Le2,Le3}.
In the local coordinates $\pi^i_{(j)} = \frac{S^i}{S^j}$
one can build the infinitesimal generators of the
Lie algebra $AlgSU(N)$.
Then one has to use explicit form
$\Phi^i_\sigma$ for $N^2-1$ of infinitesimal generators of
the Lie algebra $AlgSU(N)$. For example for the three-level
system, algebra $SU(3)$ has 8 infinitesimal generators which
are given by the vector fields:
\begin{eqnarray}
\vec{D}_1&=&i \frac{\hbar}{2}[[1-(\pi^1)^2]\frac{\partial}{\partial \pi^1}
-\pi^1 \pi^2 \frac{\partial}{\partial \pi^2}
-[1-(\pi^{1*})^2]\frac{\partial}{\partial \pi^{1*}}
+\pi^{1*} \pi^{2*} \frac{\partial}{\partial \pi^{2*}}] ,  \cr
\vec{D}_2&=&- \frac{\hbar}{2}[[1+(\pi^1)^2]\frac{\partial}{\partial \pi^1}
+\pi^1 \pi^2 \frac{\partial}{\partial \pi^2}
+[1+(\pi^{1*})^2]\frac{\partial}{\partial \pi^{1*}}
+\pi^{1*} \pi^{2*} \frac{\partial}{\partial \pi^{2*}}] , \cr
\vec{D}_3&=&-i \hbar[\pi^1 \frac{\partial}{\partial \pi^1}+\frac{1}{2}\pi^2
\frac{\partial}{\partial \pi^2}
+\pi^{1*} \frac{\partial}{\partial \pi^{1*}}+\frac{1}{2}\pi^{2*} \frac{\partial}{\partial
\pi^{2*}}], \cr
\vec{D}_4&=& i\frac{\hbar}{2}[[1-(\pi^2)^2]\frac{\partial}{\partial \pi^2}
-\pi^1 \pi^2 \frac{\partial}{\partial \pi^1}
-[1-(\pi^{2*})^2]\frac{\partial}{\partial\pi^{2*}}
+\pi^{1*} \pi^{2*} \frac{\partial}{\partial \pi^{1*}}] , \cr
\vec{D}_5&=& -\frac{\hbar}{2}[[1+(\pi^2)^2]\frac{\partial}{\partial \pi^2}
+\pi^1 \pi^2 \frac{\partial}{\partial \pi^1}
+[1+(\pi^{2*})^2]\frac{\partial}{\partial \pi^{2*}}
+\pi^{1*} \pi^{2*} \frac{\partial}{\partial \pi^{1*}}], \cr
\vec{D}_6&=&i\frac{\hbar}{2}[\pi^2 \frac{\partial}{\partial \pi^1}
+\pi^1 \frac{\partial}{\partial \pi^2}
-\pi^{2*}\frac{\partial}{\partial \pi^{1*}}
-\pi^{1*} \frac{\partial}{\partial \pi^{2*}}] , \cr
\vec{D}_7&=&\frac{\hbar}{2}[\pi^2 \frac{\partial}{\partial \pi^1}
-\pi^1 \frac{\partial}{\partial \pi^2}
+\pi^{2*}\frac{\partial}{\partial \pi^{1*}}
-\pi^{1*} \frac{\partial}{\partial \pi^{2*}}] , \cr
\vec{D}_8&=&-\frac{3^{1/2}}{2}i\hbar[\pi^2 \frac{\partial}{\partial \pi^2}
-\pi^{2*} \frac{\partial}{\partial \pi^{2*}}].
\end{eqnarray}

Let me assume that $|G>=\sum_{a=0}^{N-1} g^a|\hbar a>$ is a ``ground
state" of some the least action problem.
Then the velocity of the ground state evolution relative ``world
time" $\tau$ is given by the formula
\begin{eqnarray}
|\Psi> \equiv |T> =\frac{d|G>}{d\tau}=\frac{\partial g^a}{\partial
\pi^i}\frac{d\pi^i}{d\tau}|\hbar a>+\frac{\partial g^a}{\partial
\pi^{*i}}\frac{d\pi^{*i}}{d\tau}|\hbar a> \cr
=|T_i>\frac{d\pi^i}{d\tau}+|T_{*i}>\frac{d\pi^{*i}}{d\tau}=H^i|T_i>+H^{*i}|T_{*i}>,
\end{eqnarray}
is the tangent vector to the evolution curve $\pi^i=\pi^i(\tau)$,
where
\begin{eqnarray}
|T_i> = \frac{\partial g^a}{\partial \pi^i}|\hbar a>=T^a_i|\hbar a>,
\quad |T_{*i}> = \frac{\partial g^a}{\partial
\pi^{*i}}|\hbar a>=T^a_{*i}|\hbar a>.
\end{eqnarray}
Then the variation velocity of the $|\Psi>$ is given by the equation
\begin{eqnarray}\label{43}
|A> &=&\frac{d|\Psi>}{d\tau} \cr &=&
(B_{ik}H^i\frac{d\pi^k}{d\tau}+B_{ik^*}H^i\frac{d\pi^{k*}}{d\tau}
+B_{i^*k}H^{i^*}\frac{d\pi^k}{d\tau} +B_{i^*
k^*}H^{i^*}\frac{d\pi^{k*}}{d\tau})|N>\cr &+&
(\frac{dH^s}{d\tau}+\Gamma_{ik}^s
H^i\frac{d\pi^k}{d\tau})|T_s>+(\frac{dH^{s*}}{d\tau}+\Gamma_{i^*k^*}^{s*}
H^{i*}\frac{d\pi^{k*}}{d\tau})|T_{s*}>,
\end{eqnarray}
where I introduce the matrix $\tilde{B}$ of the second quadratic
form whose components are defined by following equations
\begin{eqnarray}\label{45}
B_{ik}|N> =\frac{\partial |T_i>}{\partial \pi^k}-\Gamma_{ik}^s|T_s>,
\quad B_{ik^*}|N> = \frac{\partial |T_i>}{\partial \pi^{k*}} \cr
B_{i^*k}|N> =\frac{\partial |T_{i*}>}{\partial \pi^k}, \quad B_{i^*
k^*}|N> = \frac{\partial |T_{i*}>}{\partial
\pi^{k*}}-\Gamma_{i^*k^*}^{s*}|T_{s*}>
\end{eqnarray}
through the state $|N>$ normal to the ``hypersurface'' of the ground
states. I should emphasize that ``world time" is the  time of evolution from
the one GCS to another one which is physically distinguishable.
Thereby the unitary evolution of the action amplitudes generated by
leads in general to the non-unitary evolution of the tangent
vector to $CP(N-1)$ associated with ``state vector" $|\Psi>$.
Assuming that the ``acceleration'' $|A>$ is gotten by the
action of some linear ``Hamiltonian" $\hat{L}$ describing the
evolution (or a measurement), one has the ``Schr\"odinger equation
of evolution"
\begin{eqnarray}\label{56}
\frac{d|\Psi>}{d\tau}&=&-i\hat{L}|\Psi> \cr
&=&(B_{ik}H^i\frac{d\pi^k}{d\tau}+B_{ik^*}H^i\frac{d\pi^{k*}}{d\tau}
+B_{i^*k}H^{i^*}\frac{d\pi^k}{d\tau} +B_{i^*
k^*}H^{i^*}\frac{d\pi^{k*}}{d\tau})|N> \cr &+&
(\frac{dH^s}{d\tau}+\Gamma_{ik}^s
H^i\frac{d\pi^k}{d\tau})|T_s>+(\frac{dH^{s*}}{d\tau}+\Gamma_{i^*k^*}^{s*}
H^{i*}\frac{d\pi^{k*}}{d\tau})|T_{s*}>.
\end{eqnarray}
This ``Hamiltonian" $\hat{L}$ is non-Hermitian and its expectation
values is as follows:
\begin{eqnarray}\label{57}
<N|\hat{L}|\Psi>&=&
i(B_{ik}H^i\frac{d\pi^k}{d\tau}+B_{ik^*}H^i\frac{d\pi^{k*}}{d\tau}
+B_{i^*k}H^{i^*}\frac{d\pi^k}{d\tau} +B_{i^*
k^*}H^{i^*}\frac{d\pi^{k*}}{d\tau}),\cr <\Psi|\hat{L}|\Psi>&=&
iG_{p^*s}(\frac{dH^s}{d\tau}+\Gamma_{ik}^s
H^i\frac{d\pi^k}{d\tau})H^{p*}+iG_{ps^*}(\frac{dH^{s*}}{d\tau}+\Gamma_{i^*
k^*}^{s*} H^{i^*}\frac{d\pi^{k*}}{d\tau})H^p\cr
&=&i<\Psi|\frac{d}{d\tau}|\Psi>.
\end{eqnarray}
The minimization of the $|A>$ under the transition from point $\tau$
to $\tau+d\tau$ may be achieved by the annihilation of the
tangential component
\begin{equation}
\frac{dH^s}{d\tau}+\Gamma_{ik}^s H^i\frac{d\pi^k}{d\tau}=0, \quad
\frac{dH^{s*}}{d\tau}+\Gamma_{i^* k^*}^{s*}
H^{i^*}\frac{d\pi^{k*}}{d\tau}=0
\end{equation}
i.e. under the condition of the affine parallel transport of the
Hamiltonian vector field. The last equations in (26) shows that the
affine parallel transport of $H^i$ agrees with Fubini-Study metric
leads to Berry's ``parallel transport" of $|\Psi>$.

\section{Dynamical space-time as ``objective observer"}
I have assumed that the quantum measurement of the LDV being encoded with help
infinitesimal Lorentz transformations of qubit spinor leads to emergence of
the dynamical space-time that takes the place of the objective ``quantum
measurement machine" formalizing the process of numerical encoding the results
of comparisons of LDV's. Two these procedures are described below.

\subsection{LDV's comparison}
Local representation of unitary group $SU(N)$ is reliable geometric
tool for classification of the GCS motions in $CP(N-1)$ during
quantum dynamics due to interaction or self-interaction. This
evolution of GCS may be used in objective measuring process. Two essential
components of any measurement are identification and comparison. The Cartan's idea
of reference to the previous infinitesimally close GCS has been used. So one could
avoid the necessity of the ``second body" used as a reference frame. Thereby, LDV
is now a new important element of quantum dynamics \cite{Le4}. We should be able
to compare some LDV at two infinitesimally close GCS represented by points of $CP(N-1)$.
Since LDV's are vector fields on $CP(N-1)$, the most natural mean of comparison of
the LDV's is affine parallel transport agrees with Fubini-Study metric \cite{Le1}.
This parallel transport expresses the conservation law of local Hamiltonian
\begin{equation}\label{20}
\frac{\delta H^i}{\delta \tau}=
\frac{\delta (\Omega^{\sigma}(\tau)\Phi_{\sigma}^i) }{\delta \tau}= 0,
\end{equation}
reflecting objective identification of evolving quantum process. It gives a natural
mean for the comparison of LDV at different GCS's. Field equations will be discussed
in the paragraph 5.
\subsection{Encoding the results of comparison}
The results of the comparison of LDV's should be formalized by numerical encoding.
Thus one may say that ``LDV has been measured". The invariant encoding is based on
the geometry of $CP(N-1)$ and LDV dynamics, say, dynamics of the local Hamiltonian field.
Its affine parallel transport expresses the self-conservation of quantum object
associated with ``particle" or ``unparticle". In order to build the qubit spinor 
$\eta$ of the quantum question
$\hat{Q}$ \cite{Le5} two orthogonal vectors $\{|N>,|\Psi>\}$ have been used.
Here $|N>$ is the complex normal and $|\Psi>$ tangent vector to $CP(N-1)$.
I will use following qubit spinor 
\begin{eqnarray}\label{24}
\eta=\left(
  \begin{array}{cc}
    \eta^0_{(\pi^1,...,\pi^{N-1})}  \\
    \eta^1_{(\pi^1,...,\pi^{N-1})} \\
  \end{array}
\right) = \left(
  \begin{array}{cc}
    \frac{<N|\hat{L}|\Psi>}{<N|N>}  \\
    \frac{<\Psi|\hat{L}|\Psi>}{<\Psi|\Psi>} \\
  \end{array}
\right)
\end{eqnarray}
for the measurement of the Hamiltonian $\hat{H}$ at corresponding GCS.
\subsection{Quantum boosts and angle velocities}
Any two infinitesimally close spinors $\eta$ and $\eta+\delta
\eta$ may be formally connected with infinitesimal ``Lorentz spin transformations
matrix'' \cite{G}
\begin{eqnarray}\label{31}
\hat{L}=\left( \begin {array}{cc} 1-\frac{i}{2}\delta \tau ( \omega_3+ia_3 )
&-\frac{i}{2}\delta \tau ( \omega_1+ia_1 -i ( \omega_2+ia_2)) \cr
-\frac{i}{2}\delta \tau
 ( \omega_1+ia_1+i ( \omega_2+ia_2))
 &1-\frac{i}{2}\delta \tau( -\omega_3-ia_3)
\end {array} \right).
\end{eqnarray}
I have assumed that there is not only formal but dynamical reason for such transition
when Lorentz reference frame ``follows" for GCS.
Then ``quantum accelerations" $a_1,a_2,a_3$ and ``quantum angle velocities" $\omega_1,
\omega_2, \omega_3$ may be found in the linear approximation from
the equation $\delta \eta = \hat{L} \eta-\eta$, or, strictly speaking, from
its consequence - the equations for the velocities $\xi$ of $\eta$ spinor variations
\begin{eqnarray}
\hat{R}\left(
  \begin{array}{cc}
    \eta^0  \cr
    \eta^1
  \end{array}
\right) =
\frac{\hat{L}-\hat{1}}{\delta \tau}\left(
  \begin{array}{cc}
    \eta^0  \cr
    \eta^1
  \end{array}
\right) = \left(
  \begin{array}{cc}
    \xi^0 \cr
    \xi^1
  \end{array}
\right).
\end{eqnarray}
One should take into account that in the linear approximation
the normal component of the qubit spinor does not change, i.e. $\xi^0=0$ but tangent
component $\eta^1$ subjected the affine parallel transport back to the initial GCS:
$\xi^1=\frac{\delta \eta^1}{\delta \tau}=-\Gamma \eta^1 \frac{\delta \pi}{\delta \tau}$.
If one put $\pi=e^{-i\phi} \tan(\theta/2)$ then $\frac{\delta \pi}{\delta \tau}=
\frac{\partial \pi}{\partial \theta}\frac{\delta \theta}{\delta \tau}+
\frac{\partial \pi}{\partial \phi}\frac{\delta \phi}{\delta \tau}$, where
\begin{eqnarray}
\frac{\delta \theta}{\delta \tau}=-\omega_3\sin(\theta)-((a_2+\omega_1)\cos(\phi)+
(a_1-\omega_2)\sin(\phi))\sin(\theta/2)^2 \cr
-((a_2-\omega_1)\cos(\phi)+
(a_1+\omega_2)\sin(\phi))\cos(\theta/2)^2; \cr
\frac{\delta \phi}{\delta \tau}=a_3+(1/2)(((a_1-\omega_2)\cos(\phi)-
(a_2+\omega_1)\sin(\phi))\tan(\theta/2) \cr
-((a_1+\omega_2)\cos(\phi)-
(a_2-\omega_1)\sin(\phi))\cot(\theta/2)),
\end{eqnarray}
then one has linear non-homogeneous system of 6 real equation
\begin{eqnarray}
\Re(\hat{R}_{00}\eta^0+\hat{R}_{01}\eta^1)&=&0, \cr
\Im(\hat{R}_{00}\eta^0+\hat{R}_{01}\eta^1)&=&0, \cr
\Re(\hat{R}_{10}\eta^0+\hat{R}_{11}\eta^1+\Gamma \eta^1 \frac{\delta \pi}{\delta \tau})&=&0,
\cr
\Im(\hat{R}_{10}\eta^0+\hat{R}_{11}\eta^1+\Gamma \eta^1 \frac{\delta \pi}{\delta \tau})&=&0,
\cr
\frac{\delta \theta}{\delta \tau}&=&F_1, \cr
\quad \frac{\delta \phi}{\delta \tau}&=&F_2,
\end{eqnarray}
giving $\vec{a},\vec{\omega}$ as functions of local coordinates of GCS and 2 real
perturbation frequencies $F_1, F_2$ of coset deformation acting along some geodesic in $CP(N-1)$.
Since $CP(N-1)$ is totally geodesic manifold \cite{KN}, each geodesic belongs to some
$CP(1)$ parameterized by single $\pi$ used above.

Quantum lump takes the place of extended ``pointer".
This extended pointer may be mapped onto dynamical space-time if one assumes
that transition from one GCS to another is accompanied by dynamical
transition from one Lorentz frame to another, see Figure 2.
Thereby, infinitesimal Lorentz transformations define small
``dynamical space-time'' coordinates variations. It is convenient to take
Lorentz transformations in the following form
\begin{eqnarray}
ct'&=&ct+(\vec{x} \vec{a}) \delta \tau \cr
\vec{x'}&=&\vec{x}+ct\vec{a} \delta \tau
+(\vec{\omega} \times \vec{x}) \delta \tau
\end{eqnarray}
where I put
$\vec{a}=(a_1/c,a_2/c,a_3/c), \quad
\vec{\omega}=(\omega_1,\omega_2,\omega_3)$ \cite{G} in order to have
for $\tau$ the physical dimension of time. The expression for the
``4-velocity" $ V^{\mu}$ is as follows
\begin{equation}\label{29}
V^{\mu}=\frac{\delta x^{\mu}}{\delta \tau} = (\vec{x} \vec{a},
ct\vec{a}  +\vec{\omega} \times \vec{x}) .
\end{equation}
The coordinates $x^\mu$ of imaging point in dynamical space-time serve here merely for
the parametrization of the energy distribution in the ``field
shell'' arising under ``morphogenesis" described by quasi-linear field
equations \cite{Le2,Le6,Le7}.

\section{Morphogenesis of the lump and unparticle sectors}
The conservation law of local Hamiltonian is expressed by the
affine parallel transport  (22) in $CP(N-1)$. This parallel transport
provides the ``self-conservation" of extended object, i.e.
the affine gauge fields couple the soliton-like system \cite{Le2,Le3}.

The field equations for the $SU(N)$ parameters $\Omega^{\alpha}$
dictated by the affine parallel transport of the Hamiltonian vector field
$H^i=\hbar \Omega^{\alpha}\Phi^i_{\alpha}$ (5)
read as quasi-linear PDE together with ``riccator" describing evolution of GCS
\begin{equation}\label{40}
\frac{\delta \Omega^{\alpha}}{\delta \tau} = V^{\mu} \frac{\partial
\Omega^{\alpha}}{\partial x^{\mu} } = -
(\Gamma^m_{mn} \Phi_{\beta}^n+\frac{\partial
\Phi_{\beta}^n}{\partial \pi^n}) \Omega^{\alpha}\Omega^{\beta},
\quad \frac{d\pi^k}{d\tau}= \Phi_{\beta}^k \Omega^{\beta},
\end{equation}
comprising self-consistent system. It is impossible of course to solve this
self-consistent problem
analytically even in this simplest case of the two state system, but
it is reasonable to develop a numerical approximation in the
vicinity of the following exact solution.

Let me discuss initially only quasi-linear PDE obtained as a consequence of the
parallel transport of the local Hamiltonian
\begin{equation}
 (\vec{x} \vec{a},
ct\vec{a}  +\vec{\omega} \times \vec{x}) \frac{\partial
\Omega^{\alpha}}{\partial x^{\mu} } = -
(\Gamma^m_{mn} \Phi_{\beta}^n+\frac{\partial
\Phi_{\beta}^n}{\partial \pi^n}) \Omega^{\alpha}\Omega^{\beta}
\end{equation}
for two-level system living in $CP(1)$  \cite{Le2,Le6,Le7}. In this simplest case of
GCS dynamics with coordinate $\pi=u+iv$ the indexes are as follows:
$1\leq \alpha,\beta \leq3,\quad i,k,n=1$, and the field components
$\Omega^1=(\omega+i\gamma) \sin \Theta \cos \Phi$,
$\Omega^2=(\omega+i\gamma) \sin \Theta \sin \Phi$,
$\Omega^3=(\omega+i\gamma) \cos \Theta $ that should be defined. 
This system in the case of the spherical symmetry being splited into the
real and imaginary parts takes the form
\begin{eqnarray}
\matrix{ (r/c)\omega_t+ct\omega_r=-2\omega \gamma F(u,v), \cr
(r/c)\gamma_t+ct\gamma_r=(\omega^2 - \gamma^2) F(u,v), \cr u_t=\kappa
U(u,v,\omega,\gamma), \cr v_t=\kappa V(u,v,\omega,\gamma), }
\label{self_sys}
\end{eqnarray}
where
$\kappa$ is a coefficient and $U(u,v,\omega,\gamma), V(u,v,\omega,\gamma)$ are functions
which I avoid to write explicitly here.
Let me put $\omega=\rho
\cos \psi, \quad \gamma=\rho \sin \psi$, then, assuming for
simplicity that $\omega^2+\gamma^2=\rho^2=constant$, the two first
PDE's may be rewritten as follows:
\begin{equation}
\frac{r}{c}\psi_t+ct\psi_r=F(u,v) \rho \cos \psi.
\end{equation}
The two exact solutions of this quasi-linear PDE is as follows
\begin{eqnarray}
\psi_{1}(t,r)= \arctan \frac{\exp(2c\rho F(u,v)
f(r^2-c^2t^2))(ct+r)^{2\rho F(u,v)}-1}{\exp(2c\rho F(u,v)
f(r^2-c^2t^2))(ct+r)^{2\rho F(u,v)}+1},
\end{eqnarray}
and
\begin{eqnarray}
\psi_{2}(t,r)= \arctan \frac{2 \exp(c\rho F(u,v)
f(r^2-c^2t^2))(ct+r)^{\rho F(u,v)}}{\exp(2c\rho F(u,v)
f(r^2-c^2t^2))(ct+r)^{2\rho F(u,v)}+1},
\end{eqnarray}
where $f(r^2-c^2t^2)$ is an arbitrary function of the interval.
What is the physical interpretation of these solutions may be given?
It is interesting that this non-monotonic distribution of the force field
$\psi_{1}(t,r)$ describing ``lump" \cite{Le1,Le2,Le6,Le7} that looks like a bubble
in the dynamical space-time. These field
equations describes energy distribution in the lump which does not exist
a priori but is becoming during the self-interaction, see Figure 3.
\begin{figure}[h]
  \includegraphics[width=2in]{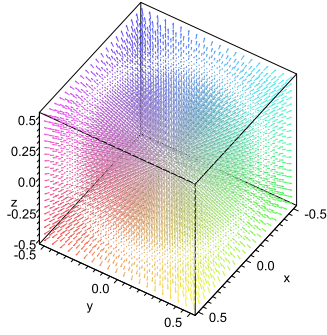}\\
  \caption{The non-monotonic distribution of the force field
in the lump looks like a bubble in the dynamical space-time.}\label{fig.3}
\end{figure}
It should be noted that attempts to treat the field dynamics literally in
spirit of ``particle in potential" are almost hopeless since we have self-consistent
dynamics. The monotonic solution $\psi_{2}(t,r)$ looks like unparticle entity 
corresponding to imaginary field mass $i\omega(r,t)$.

In order to realize the physical interpretation of these equations I will
find the stationary solution for (32). Let me put $\xi=r-ct$. Then
one will get ordinary differential equation
\begin{equation}
\frac{d\Psi(\xi)}{d \xi} = -F(u,v) \rho \frac{\cos \Psi(\xi)}{\xi}.
\end{equation}
Two solutions
\begin{equation}
\Psi(\xi) =arctan(\frac{\xi^{-2M} e^{-2CM}-1}{\xi^{-2M} e^{-2CM}+1},
\frac{2\xi^{-M} e^{-2CM}}{\xi^{-2M} e^{-2CM}-1}),
\end{equation}
where $M=F(u,v) \rho$ are concentrated in the vicinity of the
light-cone looks like solitary waves, see Fig.4.
\begin{figure}[h]
  \includegraphics[width=4in]{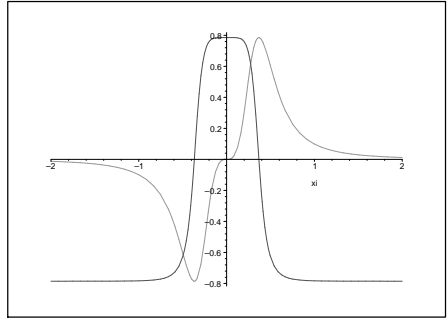}\\
  \caption{Two solutions of (35) in the light-cone vicinity.}\label{fig.4}
\end{figure}

The problem of the physical status of these field equation may be solved
if one could point out some transformations from found equations to well known
relativistic, say, Dirac equation. I almost sure that it is impossible
to find this transition as perturbation in small parameter.
Here I will give only some hints for such transition.

Standard Dirac's equation
\begin{equation}
\hat{\gamma}^{\mu} \frac{\partial \psi}{\partial x^{\mu}}+\frac{imc}{\hbar}
\hat{\gamma}^5\psi=0
\end{equation}
is linear. Dirac assumed that matrices $\hat{\gamma}^{\mu}$ should be
coordinate independent since empty Minkowskian space-time is homogeneous and isotropic.
However Dirac's equation in curved space-time should have coordinate-dependent
matrices $\hat{\gamma}^{\mu}(x) \equiv b^{\mu}_a(x) \hat{\gamma}^a$,
where $b^{\mu}_a(x)$ is vierbein defined as follows:
$g^{\mu\nu}=b^{\mu}_a(x)b^{\mu}_a(x)\eta^{ab}$ \cite{Parker}.
It is known that matrices $\hat{\gamma}^{\mu}$ have a sense of instant
velocities with modulus $c$.
Quasi-linear equation (30) has similar structure but ``4-velocity"
$V^{\mu}=\frac{\delta x^{\mu}}{\delta \tau} = (\vec{x} \vec{a},
ct\vec{a}  +\vec{\omega} \times \vec{x})$ of imaging point evidently depends on
coordinates in DST. They serves as parameters of field distribution in the lump and 
unparticle energy distribution. Probably it is possible to establish some relations
between $V^{\mu}(x)$ and $\hat{\gamma}^{\mu}(x)$ but presently this connection is 
unclear. 

\section{Conclusion}
1. Action states serve as ``initial" and ``final" conditions in fixed setup.
Manipulations with quantum amplitudes shows that two setups $S_1$ and $S_2$
generates generally
two different amplitudes $|S_1>$ and $|S_2>$ of outcome events.

2. It is reasonable to find physical invariance lurked behind $|S_1>$ and $|S_2>$.
``Relativity to measuring device``
by Fock and ``functional relativity" by A. Kryukov express this invariance
in the global manner. Since there is no strict definition of the
measuring device in terms of standard quantum theory, the special ad hoc example
of manipulation with improper states like plane wave and $\delta$-function have been used
in order to show functional invariance of state equation before and after measurement.

3. I use flexible setup for transition to local quantum reference frame in super-relativity.
Amplitude of outcome event $|S>$, its GCS and dynamical group $SU(N)$ with $N^2-1$
non-Abelian fields parameters $\Omega^{\alpha}$ are main ingredients for objective quantum
measurement. Desirable quantum localization is realized in functional space:
infinitesimal variation of fields parameters $\Omega^{\alpha}$ defines local dynamical
variables (LDV) expressed in local coordinates $\pi^1,...,\pi^{N-1}$. Non-linear
realization of  $SU(N)$ generators by tangent
vector field to $CP(N-1)$ serves for invariant classification of quantum motions
and particle/unparticle excitations instead of classification of ``elementary" 
quantum particles.

4. Objective quantum measurement of LDV creates dynamical space-time due to:

a. Comparison of LDV in infinitesimally close GCS is provided by non-Abelian affine gauge
field agreed with Fubini-Study metric,

b. Qubit spinor encoding of the result of this comparison whose components are
parameterized by quantum boosts and quantum rotations that define dynamics of attached
local Lorentz frame in DST.

5. Identification of quantum objects (processes) and its conservation law expresses
by parallel transport of local Hamiltonian. Quasi-linear PDE is consequence of
this conservation law that generate dynamics of GCS and morphogenesis of ``field shell".
Particle and unparticle sectors of these excitations should be classified by comparison with
know quantum field equations.

\section{Discussion}
1. The intrinsically geometric scheme of the quantum
measurement of local dynamical variable has been proposed.
The self-interaction supporting localizable ``lump" configuration
arose due to the breakdown of global $G=SU(N)$ symmetry is
used for such measurement and it is represented by the affine gauge
``field shell" propagated in the dynamical state-dependent
space-time.

2. The concept of ``super-relativity" \cite{Le1,Le2} is in fact a
different kind of attempts of ``hybridization" of internal and
space-time symmetries. In distinguish from SUSY where a priori
exists the extended space-time - ``super-space", in my approach the
dynamical space-time arises under ``yes/no" quantum measurement of
$SU(N)$ local dynamical variables.

3. The locality in the quantum phase space $CP(N-1)$ leads to
extended quantum particles - ``field shell" that obey the
quasi-linear PDE \cite{Le2}.

4. The main technical problem
is to find non-Abelian gauge field arising from conservation law of the local Hamiltonian
vector field. The last one may be expressed as parallel transport of local
Hamiltonian in projective Hilbert space $CP(N-1)$. Co-movable local ``Lorentz frame"
being attached to GCS is used for qubit encoding result of comparison of the
parallel transported local Hamiltonian in infinitesimally close points. This
leads to quasi-linear relativistic field equations with soliton-like solutions
for ``field shell" in emerged DST. The terms ``comparison" and ``encoding" resemble
human's procedure, but here they have objective content realized in invariant
quantum dynamics.

There is a possibility for generalization of scalar in DST ``field shell"
$\Omega^{\sigma}\Phi_{\sigma}^i$ to vector $\Omega_{\mu}^{\sigma}\Phi_{\sigma}^i$ and
tensor fields  $\Omega_{\mu \nu}^{\sigma}\Phi_{\sigma}^i$ assuming invariant contraction
in iso-index $\sigma$. Then will arise more complicated field equation with essential
dependence of global space-time structure since one need to know metric
connection $\Gamma^{\lambda}_{\mu \nu}$ for covariant derivatives.

5. One need to find connection between quasi-linear field equations and known filed equation
(like Dirac equation). Probably it is possible to use some
analogy with Skyrmion field quantization \cite{Aitchison} although there is of
course essential difference between lump and monopole solutions.

6. DST forms granular structure of global space-time and paves a way to build 
quantum gravity ``from inside".


\vskip 0.2cm
\begin{thebibliography}{99}
\bibitem{Planck1}
M. Planck, Ann. Phys., {\bf 4}, 553 (1901).
\bibitem{Einstein_Q}
A. Einstein, Ann. Phys., {\bf 17}, 132 (1905).
\bibitem{dB}
L. de Broglie, Recherches sur la Th\'eorie des Quanta, (Ann. de Phys.
$10^e$ s\'erie, t.III (Janvier-F\'evrier 1925). Translated by A.F. Kracklauer,
\copyright AFK, 2004.
\bibitem{Sch1}
E. Schr\"odinger, Ann. Phys. {\bf 79}, 361, 1926.
\bibitem{Dirac1}
P.A. Dirac, Proc. Royal. Soc. A {\bf 114}, 243 (1927).
\bibitem{Schroer}
B. Schroer, 0805.1911v2 [hep-th].
\bibitem{EPR}
A. Einstein, B. Podolsky and N. Rosen, Phys.Rev. {\bf 47}, 777 (1935).
\bibitem{Schr}
E. Schr\"odinger, Naturewissenschafen 23: pp.807-812; 823-828; 844-849 (1935).
\bibitem{Sch2}
E. Schr\"odinger, Nanurwissenschaften, {\bf 14}, H 28, 664, 1926.
\bibitem{Feynman1}
R.P. Feynman, \emph{QED The strange theory of light and matter}, Princeton,
New Jersey: Princeton University Press (1985).
\bibitem{G}
C.W. Misner, K.S. Thorne, J.A. Wheeler,{\it Gravitation}, W.H.Freeman
and Company, San Francisco, 1973.
\bibitem{Kryukov1}
A.A. Kryukov, Found. Phys. {\bf 36}, 175 (2006).
\bibitem{Kryukov2}
A.A. Kryukov, Found. Phys. {\bf 34}, 1225 (2004).
\bibitem{KN}
S. Kobayashi and K. Nomizu, {\it Foundations of Differential Geometry, V. II},
Interscience Publishers, New York-London-Sydney, 1969.
\bibitem{Le1}
P. Leifer, Found. Phys. {\bf 27}, (2) 261 (1997).
\bibitem{Le2}
P. Leifer, Annales de la Fondation Louis de Broglie, {\bf 32}, (1) 25
(2007).
\bibitem{Le6}
P. Leifer, arXiv:gr-qc/0503083.
\bibitem{Le7}
P. Leifer, L.P. Horwitz, arXiv:gr-qc/0505051 v2.
\bibitem{Georgi}
H. Georgi, arXiv:hep-th/0703260v3.
\bibitem{Yuan}
K. Cheung, W.-Y. Keung and T.-C. Yuan, arXiv:0704.2588v3 [hep-th].
\bibitem{Le3}
P. Leifer, Found.Phys.Lett., {\bf 18}, (2) 195 (2005).
\bibitem{Bl1}
D.I. Blochntzev, ``Uspechy Phys. Nauk", {\bf XLIV}, No.1, 104 (1951).
\bibitem{Bl2}
D.I. Blochntzev, ``Uspechy Phys. Nauk", {\bf XLII}, No.1, 76 (1950).
\bibitem{Fock}
V.A. Fock, \emph{Basis of quantum mechanics}, Nauka, Moskow (1976).
\bibitem{Le4}
P. Leifer, JETP Letters, {\bf 80}, (5) 367 (2004).
\bibitem{Penrose}
R. Penrose, {\it The Road to Reality}, Alfred A.Knopf, New-York,
(2005).
\bibitem{Einstein1}
A. Einstein, Ann. Phys. {\bf 17}, 891 (1905).
\bibitem{Einstein2}
A. Einstein, Ann. Phys. {\bf 49}, 769 (1916).
\bibitem{Le5}
P. Leifer, arXiv:0808.3172v1 [physics.gen-ph].
\bibitem{Le8}
P. Leifer, arXiv:0812.0065v1 [physics.gen-ph].
\bibitem{Parker}
X. Huang and L. Parker, arXiv:0811.2296v1 [hep-th].
\bibitem{Aitchison}
I.J.R. Aitchison, Acta Physica Polonica, Vol. B18, No.3, 207 (1987).
\end{thebibliography}
\end{document}